\documentclass[%
 reprint,
 amsmath,amssymb,
 aps,prc
]{revtex4-2}
\usepackage[colorlinks, citecolor=red]{hyperref}
\usepackage{graphicx}
\usepackage{epstopdf}
\usepackage{dcolumn}
\usepackage{bm}
\usepackage{natbib}
\usepackage{multirow}
\begin{document}
\preprint{APS/123-QED}
\title{Kaon-meson condensation and  \texorpdfstring{$\Delta$}{}  resonance in hyperonic stellar matter within a relativistic mean-field model}

\author{Fu Ma$^{1}$, Chen Wu$^{2}$\footnote{Electronic address: wuchenoffd@gmail.com} and Wenjun Guo$^{1}$ } \affiliation{
\small 1. University of Shanghai for Science and Technology, Shanghai 200093, China\\
\small 2. Shanghai Advanced Research Institute, Chinese Academy of Sciences, Shanghai 201210, China}

\begin{abstract}
We study the equation of state of dense baryon matter within the
relativistic mean-field model, and we include ${\Delta}$(1232) isobars into IUFSU model with hyperons and consider the possibility of kaon meson condensation. We find that it is necessary to consider the $\Delta$ resonance state inside the massive neutron star. The critical density of Kaon mesons and hyperons is shifted to a higher density region, in this respect an early appearance of {$\Delta$} resonances is crucial to guarantee the stability of the branch of hyperonized star with the difference of the coupling parameter $x_{\sigma \Delta}$ constrained based on the QCD rules in nuclear matter. The $\Delta$ resonance produces a softer equation of state  in the low density region, which makes the tidal deformability and radius consistent with the observation of GW170817. As the addition of new degrees of freedom will lead to a softening of the equation of state, the ${\sigma}$-cut scheme, which states the decrease of neutron star mass can be lowered if one assumes a limited decrease of the ${\sigma}$-meson strength at ${\rho_B}$($\rho_B > \rho_0$), finally we get a maximum mass neutron star with $\Delta$ resonance heavier than 2$M_{\odot}$.
\end{abstract}

\maketitle

\section{\label{sec:level1}INTRODUCTION}
Astronomical observations and gravitational wave data over the past decade have placed a series of constraints on a range of properties of neutron stars(mass, radius, deformabilities, e.g). The massive neutron stars (NSs) observed, e.g., PSR J1614-2230 with $M=1.908\pm0.016M_{\odot}$ \cite{Demorest:2010bx,NANOGrav:2017wvv,Fonseca:2016tux,Ozel:2010bz},have established strong constraints on the equation of state (EOS) of nuclear matter. PSR J0348+0432 with $M=2.01\pm0.04M_{\odot}$ \cite{antoniadis2013massive}, MSP J0740+6620, with $M=2.08^{+0.07}_{-0.07}M_{\odot}$ \cite{fonseca2021refined,NANOGrav:2019jur}, and radius $12.39^{+1.30}_{-0.98}$ km obtained from NICER data \cite{riley2021nicer}. The recent observation of gravitational waves from the binary neutron star merger event GW170817 suggests that the dimensionless combined tidal deformability ${\Lambda}$ is considered to be less than 720 at 90\%
confidence level based on low spin priors \cite{LIGOScientific:2018hze}, while a lower limit with  ${\Lambda \geq}$ 197 is obtained based on electromagnetic observations of the transient counterpart AT2017gfo \cite{Coughlin:2018miv}. These astronomical observations constrain the tidal deformability of a 1.4 $M_{\odot}$ mass neutron star and thus strong interactions in dense nuclear matter. These upper limits indicate that the  EOS of stellar material is softened at this (intermediate) density. One way to solve this problem is to introduce new degrees of freedom(hyperons \cite{Schaffner:1995th,Wu:2011zzb}, ${\Delta}$ resonance \cite{Sun:2018tmw,Maslov:2017lkx,Kolomeitsev:2016ptu,Sedrakian:2022kgj,Drago:2014oja,Glendenning:1984jr,Dexheimer:2021sxs,Marquez:2022fzh,Schurhoff_2010}, kaon meson condensation \cite{Thapa:2020usm,Maruyama:2005tb,Brown:2005yx,Shao:2010zz,Char:2014cja}), as the density of nucleons increases, the appearance of hyperons, ${\Delta}$ resonance, and kaon condensation inevitably softens the equation of state, resulting in a neutron star with a mass radius consistent with astronomical observations.
\par{}
As the density of nucleons increases, hadron degrees of freedom inside the neutron star are excited into strangeness-bearing hyperons, they affect the stellar structure and evolution in various ways \cite{Burgio:2021vgk,Logoteta:2021iuy}. Although the existence of hyperons inside neutron stars is inevitable, its appearance will significantly lead to a softening of the equation of state, resulting in a decrease in the maximum mass of the neutron star, which does not correspond to the observation of massive neutron stars ($2M_\odot$), which is known as hyperons puzzle \cite{burgio2011hyperon,Lonardoni:2014bwa,Bombaci:2016xzl}. In order to guarantee a stiffen EOS and massive neutron stars, density covariant functional theory has been chosen to study neutron stars containing hyperons \cite{bonanno2012composition,Colucci:2013pya,Fortin:2016hny,Chen:2007kxa,Drago:2014oja,Li:2019fqe,Li:2019fqe,li2019implications}. However, with the constraints on tidal deformation and radii imposed by astronomical observations, the application of this theoretical model is subject to some limitations \cite{Zhu:2018ona,Malik:2018zcf}.
\par{}
Although there are many speculations about the existence of hyperons inside neutron stars, there is little discussion about the ${\Delta}$ resonance. One reason is that early work suggested that the critical densities of ${\Delta}$ resonances in the RMF model with the same strength of the meson field as the nucleon case have exceeded the densities of the core of typical neutron stars \cite{Glendenning:1984jr,Glendenning:1991es}, which is considered out of the realm of astrophysics, and another reason is that the occurrence of ${\Delta}$ resonance leads to a softening of the equation of state, which has become a ${\Delta}$ puzzle \cite{Drago:2014oja} in some literature, the same as the hyperon puzzle. However, recent work has shown that considering the ${\Delta}$ resonance inside a neutron star reduces the radius with a standard mass of 1.4$M_{\odot}$ NS, and that the equation of state does not change significantly \cite{Li:2018qaw,li2019implications,Ribes:2019kno}.
\par{}
Another new degree of freedom for non-nucleons in dense stars includes various mesons(kaon, pion) condensates\cite{Glendenning:1997ak,Glendenning:1998zx}. Kaplan and Nelson have suggested that the ground state of hadronic matter might form a negatively charged Kaon Bose-Einstein condensation above a certain critical density \cite{Kaplan:1986yq,Nelson:1987dg}. In the interior of a neutron star, as the density of neutrons increases, the electronic chemical potential will increase to keep the matter in $\beta$-equilibrium. When the electronic chemical potential exceeds the mass of muons, muons appear. And when the vacuum mass of the meson (pion, Kaon) is exceeded, as the density increases, negatively charged mesons begin to appear, which helps to maintain electrical neutrality. However, the $s$ - wave $\pi N$ scattering potential repels the ground state mass of the $\pi$ meson and prevents the generation of the $\pi$ meson \cite{Glendenning:1984jr}. With the increase of density, the energy $\omega_{K}$ of a test Kaon in the pure normal phase can be computed as a function of the nucleon density. The Kaon energy will decrease while the chemical potential of Kaon increases with the density. When the condition $\omega_{K}=\mu_{e}$ is achieved, the Kaon will occupy a small fraction of the total volume, then $K^{-}$ will be more advantageous than electrons as a neutralizer for positive charges, and this will open the possibility of the appearance of Kaon condensates.
\par{}
Although many scholars have previously proposed various density covariant functional theories or realistic nuclear potential in order to obtain more massive neutron stars containing hyperons.  However these theories are usually used to consider neutrons, protons and leptons ($n, p, e, {\mu^{-}}$) because of the hyperon puzzle, and considering hyperons in the RMF framework leads to a reduction of the maximum mass of neutron stars and thus does not satisfy astronomical observations. However, the ${\sigma} \mbox{-}$cut scheme \cite{Maslov:2015lma}, which point out that in the small scale range where the density ${\rho_B}$\textgreater${\rho_0}$, a sharp decrease in the strength of the ${\sigma}$ meson reduces the decrease in the effective mass of the nucleon, which eventually stiffens the EOS and still yields neutron stars of more than 2$M_\odot$ after considering the hyperon degrees of freedom \cite{Ma:2022fmu,Wu:2020yqp}. In this article, we use the IUFSU model \cite{Fattoyev:2010mx,Grill:2014aea} to study NS matter including hyperons, ${\Delta}$ resonance and Kaon condensates with $\sigma \mbox{-}$cut scheme.
\par
This paper is organized as follows. First, the theoretical framework is presented. Then we will study the effects of Kaon meson condensation and ${\Delta}$ resonance contains hyperons with the $\sigma$-cut scheme. Finally, some conclusions are provided.

\section{\label{sec:level2}theoretical framework}

\begin{table}.
\caption{\label{tab:Table 1.}
Strangeness mass $M$, third component of isospin $\tau_{3}$, charge $q$, total angular momentum and parity $J^{P}$ for $\Lambda^{0},\Sigma^{+,0,-}$ and $\Xi^{-,0}$ hyperons and $\Delta$ baryons.
}
\begin{ruledtabular}
\begin{tabular}{l|cccc}
\textrm{} &
\textrm{$M(MeV)$} &
\textrm{$\tau_{3}$} &
\textrm{$q(e)$} & 
\textrm{$J^{p}$} \\
$\Lambda^{0}$ & 1116 & 0 & 0 & ${(1/2)}^{+}$ \\ \hline
$\Sigma^{+}$ & 1193 & 1 & +1 & ${(1/2)}^{+}$ \\
$\Sigma^{0}$ & 1193 & 0 & 0 & ${(1/2)}^{+}$ \\
$\Sigma^{-}$ & 1193 & -1 & -1 & ${(1/2)}^{+}$ \\ \hline
$\Xi^{0}$ & 1318 & ${(1/2)}$ & 0 & ${(1/2)}^{+}$ \\
$\Xi^{-}$ & 1318 & ${(-1/2)}$ & -1 & ${(1/2)}^{+}$ \\ \hline
$\Delta^{++}$ & 1232 & ${(+3/2)}$ & +2 & ${(3/2)}^{+}$ \\
$\Delta^{+}$ & 1232 & ${(+1/2)}$ & +1 & ${(3/2)}^{+}$ \\
$\Delta^{0}$ & 1232 & ${(-1/2)}$ & 0 & ${(3/2)}^{+}$ \\
$\Delta^{-}$ & 1232 & ${(-3/2)}$ & -1 & ${(3/2)}^{+}$ \\
\end{tabular}
\end{ruledtabular}
\end{table}

In this section, we introduce the IUFSU model to study the properties of the ${\Delta}$ resonance and phase transition from hadronic to Kaon condensed matter. For the baryons matter we have considered nucleons ($n$ and $p$), and hyperons ($\Lambda, \Sigma$ and $\Xi$), ${\Delta}$ resonance (${\Delta^{++}, \Delta^{+}, \Delta^{0}, \Delta^{-}}$), kaon (${K^-}$). The exchanged mesons include the isoscalar scalar meson ($\sigma$), the isoscalar vector meson ($\omega$), the isovector vector meson ($\rho$), and strange vector meson (${\phi}$), the starting point of the extended IUFSU model is the Lagrangian density:
\begin{widetext}
\begin{equation}
\begin{aligned}
\mathcal{L}=&\sum_{B}\bar{\psi}_{B}[\emph{$i$}\gamma^{\mu}\partial{\mu}-m_B+g_{\sigma B}\sigma-g_{\omega B}\gamma^{\mu}\omega_{\mu}-g_{\phi B}\gamma^{\mu}\phi_{\mu}-{g_{\rho B}}\gamma^{\mu}\vec{\tau}\cdot\vec{\rho^{\mu}}]\psi_{B}+\\
&\sum_{D}\bar{\psi}_{D}[\emph{$i$}\gamma^{\mu}\partial{\mu}-m_D+g_{\sigma D}\sigma-g_{\omega D}\gamma^{\mu}\omega_{\mu}-g_{\phi D}\gamma^{\mu}\phi_{\mu}-{g_{\rho D}}\gamma^{\mu}\vec{\tau}\cdot\vec{\rho^{\mu}}]\psi_{D}+\\
&\frac{1}{2}\partial_{\mu}\sigma\partial^{\mu}\sigma-\frac{1}{2}m_{\sigma}^{2}\sigma^{2}-\frac{\kappa}{3!}(g_{\sigma N}\sigma)^3-\frac{\lambda}{4!}(g_{\sigma N}\sigma)^{4}-\frac{1}{4}F_{\mu \nu}F^{\mu \nu}+\frac{1}{2}m_{\omega}^{2}{\omega_{\mu}}\omega^{\mu}-\\
&\frac{1}{4}{\Phi}_{\mu \nu}{\Phi}^{\mu \nu}+\frac{1}{2}m_{\phi}^{2}{\phi_{\mu}}\phi^{\mu}+\frac{\xi}{4!}(g_{\omega N}^{2}\omega_{\mu}\omega^{\mu})^{2}+\frac{1}{2}m_{\rho}^{2}\vec{\rho}_{\mu}\cdot\vec{\rho}^{\mu}-\frac{1}{4}\vec{G}_{\mu \nu}\vec{G}^{\mu \nu}+\\
&\Lambda_{\nu}(g_{\rho N}^{2}\vec{\rho}_{\mu}\cdot\vec{\rho}^{\mu})(g_{\omega N}^{2}\omega_{\mu}\omega^{\mu})+\sum_{l} \bar{\psi}_{l}[i{\gamma}^{\mu}\partial{\mu}-m_{l}]{\psi}_{l},
\label{eq:one}
\end{aligned}
\end{equation}
\end{widetext}

with the field tensors
\begin{equation}
\begin{aligned}
&\emph{F}_{\mu \nu}=\partial_{\mu}\omega_{\nu}-\partial_{\nu}\omega_{\mu}\\
&{\Phi_{\mu \nu}}=\partial_{\mu}\phi_\nu-\partial_{\nu}\phi_{\mu}\\
&\vec{G}_{\mu \nu}=\partial_{\mu}\rho_{\nu}-\partial_{\nu}\rho_{\mu},
\end{aligned}
\end{equation}

The model contains following quantities, the baryon octet and two leptons($\emph{p,n,e},\mu,\Lambda^{0},\Sigma^{+},\Sigma^{0},\Sigma^{-},\Xi^{0},\Xi^{-}$), $\Delta$ resonances (${\Delta^{++}, \Delta^{+}, \Delta^{0}, \Delta^{-}}$), isoscalar-scalar $\sigma$, isoscalar-vector $\omega, \phi$ and isoscalar-vector $\rho$ with the masses and coupling constants. The isospin operator for the isovector-vector meson fields is represented by $\vec{\tau}$. where $\Lambda_{\nu}$ is introduced to modify the density dependence of symmetry energy. The isoscalar meson self-interactions (via $\kappa$, $\lambda$ and $\xi$ terms) are necessary for the appropriate EOS of symmetric nuclear matter, respectively. In RMF models, the operators of meson fields are replaced by their expectation values by the mean field approximation. In Table \ref{tab:Table 1.}, we list properties and coupling constants
for baryons other than nucleons in Eq.(\ref{eq:one}).

We take the Lagrangian of Kaon condensation as the same that is Ref. \cite{Glendenning:1997ak} and \cite{Glendenning:1998zx}, which reads
\begin{equation}
{\cal{L}}_{K}=D_{\mu}^{*}K^{*}D^{\mu}K-m_{K}^{* 2}K^{*}K,
\label{eq:two}
\end{equation}
where $\emph{D}_{\mu}=\partial_{\mu}+ig_{\omega K}\omega_{\mu}+ig_{\phi K}{\phi}+i\frac{g_{\rho K}}{2}\tau_{\emph{K}}\cdot\rho_{\mu}$ is the covariant derivative and the Kaon effective mass is defined as
$\emph{m}_{\emph{K}}^{*}=\emph{m}_{\emph{K}}-g_{\sigma K}\sigma$.
\par{}
Finally, with the Euler-Lagrange equation, the
equations of motion for baryons and mesons are obtained:
\begin{widetext}
\begin{equation}
\begin{aligned}
&m_{\sigma}^{2}\sigma+\frac{1}{2}{\kappa}g_{\sigma N}^{3}{\sigma}^{2}+\frac{1}{6}{\lambda}g_{\sigma N}^{4}{\sigma}^{3}=\sum_{B}g_{\sigma B}{\rho}_{B}^{S}+\sum_{D}g_{\sigma D}{\rho}_{D}^{S}+g_{\sigma K}{\rho}_{K}\\
&m_{\omega}^{2}{\omega}+\frac{\xi}{6}g_{\omega N}^{4}{\omega}^{3}+2{\Lambda_{\nu}}g_{\rho N}^{2}g_{\omega N}^{2}{\rho}^{2}{\omega}=\sum_{B}g_{\omega B}{\rho}_{B}+\sum_{D}g_{\omega D}{\rho}_{D}-g_{\omega K}{\rho}_{K}\\
&m_{\rho}^{2}{\rho}+2{\Lambda}_{\nu}g_{\rho N}^{2}g_{\omega N}^{2}{\omega}^{2}{\rho}=\sum_{B}{g_{\rho B}}{\tau}_{3B}{\rho}_{B}+\sum_{D}{g_{\rho D}}{\tau}_{3D}{\rho}_{D}-{\frac{g_{\rho K}}{2}{\rho}_{K}}\\
&m_{\phi}^{2}{\phi}=\sum_{B}g_{\phi B}{\rho}_{B}-g_{\phi K}{\rho}_{K},
\end{aligned}
\end{equation}
\end{widetext}
where $\rho_{B(D)}$ and $\rho_{B(D)}^{S}$ are the baryon($\Delta$) density and the scalar density, which reads:
\begin{equation}
\begin{aligned}
&\rho_B=\frac{\gamma k_{fB}}{6\pi^2}\\
&\rho_B^S=\frac{\gamma M^{*}}{4\pi^2}[k_{fB}E^{*}_{fB}-M^{*2}\ln(\frac{k_{fB}+E^{*}_{fB}}{M^{*2}})]
\end{aligned}
\end{equation}
$\gamma=2$ for baryons and $\gamma=4$ for $\Delta$ resonance. where $E^{*}_{fB}=\sqrt{k_{fB}^2+M^{*2}}$.
the Kaon density
\begin{equation}
\rho_{K}=2(\omega_{K}+g_{\omega K}{\omega}+g_{\phi K}{\phi}+\frac{g_{\rho K}}{2}{\rho})K^{*}K,
\end{equation}
\par{}
Now, we are in a position to discuss the coupling parameters between  baryons (nucleons, hyperons and $\Delta$) or  $K^{-}$  and meson fields. The masses of nucleons
and mesons, and the coupling constants between nucleon and mesons in IUFSU models are tabulated in Table.\ref{tab:Table 2.}
\begin{table*}
\caption{\label{tab:Table 2.}
Parameter sets for the IUFSU model discussed in the text and the meson masses $M_{\sigma}=491.5MeV$, $M_{\omega}=786MeV$, $M_{\rho}=763MeV$.
}
\begin{ruledtabular}
\begin{tabular}{lccccccr}
\textrm{Model}&
\textrm{$g_{\sigma}$} &
\textrm{$g_{\omega}$} &
\textrm{$g_{\rho}$} &
\textrm{$\kappa$} &
\textrm{$\lambda$} &
\textrm{$\xi$} &
\textrm{$\Lambda_{\nu}$}\\
IUFSU & 9.9713 & 13.0321 & 13.5899 & 3.37685 & 0.000268 & 0.03 & 0.046\\
\end{tabular}
\end{ruledtabular}
\end{table*}

\par
For the meson-hyperon couplings, we take those in the SU(6) symmetry for the vector couplings constants:
\begin{equation}
\begin{aligned}
&g_{\rho \Lambda}=0, g_{\rho \Sigma}=2g_{\rho \Xi}=2g_{\rho N}\\
&g_{\omega \Lambda}=g_{\omega \Sigma}=2g_{\omega \Xi}=\frac{2}{3}g_{\omega N}\\
&2g_{\phi \Lambda}=2g_{\phi \Sigma}=g_{\phi \Xi}=\frac{-2\sqrt{2}}{3}g_{\omega N}
\end{aligned}
\end{equation}
While the nucleons do not couple to the strange mesons, $g_{\phi N}$=0, and mass of meson ${\phi}$ takes as $M_{\phi}$=1020MeV.
The scalar couplings are usually fixed by fitting hyperon potentials with $U_{Y}^{(N)}=g_{\omega Y}\omega_{0}-g_{\sigma Y}\sigma_{0}$, where $\sigma_{0}$ and $\omega_{0}$ are the values of the scalar and vector meson strengths at saturation density \cite{Schaffner:1993qj}. We choose the hyperon-nucleon potentials of $\Lambda$, $\Sigma$ and $\Xi$ as $U^N_{\Lambda}=-30$MeV, $U^N_{\Sigma}=30$MeV and $U^N_{\Xi}=-18$MeV \cite{Hu:2021ket,Fortin:2017cvt,AGSE885:1999erv}. Table.\ref{tab:Table 3.} provides the numerical values of the meson hyperon couplings at nuclear saturation density, where
$x_{\sigma Y}={g_{\sigma Y}}/{g_{\sigma N}}$
\begin{table}.
\caption{\label{tab:Table 3.}
scalar meson hyperon coupling constants for IUFSU.
}
\begin{ruledtabular}
\begin{tabular}{lccc}
\textrm{} &
\textrm{$\Lambda$} &
\textrm{$\Sigma$} &
\textrm{$\Xi$} \\
$x_{\sigma Y}$ & 0.615796 & 0.45219 & 0.305171
\end{tabular}
\end{ruledtabular}
\end{table}
\par
The coupling constants between the vector meson and the Kaon $\emph{g}_{\omega K},g_{\rho K}$ are determined by the meson SU(3) symmetry as $g_{\omega K}=g_{\omega N}/3,g_{\rho K}=g_{\rho N}$ \cite{Char:2014cja}, and $g_{\phi K}$=4.27 for the $\phi$ meson \cite{Banik:2001yw}. The scalar coupling constant $g_{\sigma K}$ is fixed to the optical potential of the $\emph{K}^{-}$ at saturated nuclear matter,
\begin{equation}
\label{eq3}
\emph{U}_{K}(\rho_{0})=-g_{\sigma K}\sigma(\rho_{0})-g_{\omega K}
\omega(\rho_{0}),
\end{equation}
and in this paper, we carry out our calculations with a series of optical potentials ranging from -160 MeV to -120 MeV.
The $g_{\sigma K}$ can be related to the potential of Kaon at the saturated density through Eq.(\ref{eq3}). $g_{\sigma K}$ values corresponding to several values of $U_{K}$ are listed in Table \ref{tab:Table 4.}.

\begin{table}
\caption{\label{tab:Table 4.}
$g_{\sigma K}$ determined for several $U_{K}$ values in the IUFSU model.
}
\begin{ruledtabular}
\begin{tabular}{lccc}
\textrm{$U_{K}$(MeV)} &
\textrm{-120} &
\textrm{-140} &
\textrm{-160} \\
$g_{\sigma K}$ & 0.600417 & 1.144204 & 1.68799
\end{tabular}
\end{ruledtabular}
\end{table}
\par
Because experimental data on the $\Delta$ resonance is scarce, the coupling parameters between the $\Delta$ resonances and meson fields are uncertain, so we limit ourselves to considering only the couplings with $\sigma$ meson fields, which are explored in the literature \cite{Li:1997yh,Jin:1994vw}. We assume the scalar coupling ratio $x_{\sigma \Delta}$=$g_{\sigma \Delta}$/$g_{\sigma N}$ \textgreater 1 with a value close to the mass ratio of the $\Delta$ and the nucleon \cite{Kosov:1998gp}, and adopt three different choices ($x_{\sigma \Delta}$=1.05, $x_{\sigma \Delta}$=1.1 and $x_{\sigma \Delta}$=1.15) \cite{Drago:2013fsa}, for $x_{\omega \Delta}$ and $x_{\rho \Delta}$ we take as $x_{\omega \Delta}$=$g_{\omega \Delta}$/$g_{\omega N}$=1.1 and $x_{\rho \Delta}$=$g_{\rho \Delta}$/$g_{\rho N}$=1 \cite{Thapa:2021kfo}. Similar to the nucleons, $\Delta$ resonances do not couple to meson $\phi$, so $g_{\phi \Delta}$=0.

By solving the Euler-Lagrangian equation of Kaon we obtain equation of motion: $[\emph{D}_{\mu}\emph{D}^{\mu}+m_{\emph{K}}^{* 2}]K=0$. We can then derive the dispersion relation for the Bose-condensation of $\emph{K}^{-}$, which reads
\begin{equation}
\omega_{K}=m_{K}-g_{\sigma K}\sigma-g_{\omega K}\omega-g_{\phi K}\phi-\frac{g_{\rho K}}{2}\rho,
\end{equation}
\par{}
For the neutron matter with baryons and charged leptons, the $\beta$-equilibrium conditions are guaranteed with the following relations of chemical potentials for different particles:
\begin{equation}
\begin{aligned}
&\mu_{p}=\mu_{\Sigma^{+}}=\mu_{\Delta^{+}}\\
&\mu_{\Lambda}=\mu_{\Sigma^{0}}=\mu_{\Xi^{0}}=\mu_{\Delta^{0}}=\mu_{n}\\
&\mu_{\Sigma^{-}}=\mu_{\Xi^{-}}=\mu_{\Delta^{-}}=2\mu_{n}-\mu_{p}\\
&\mu_{\Delta^{++}}=2\mu_{p}-\mu_{n}\\
&\mu_{\mu}=\mu_{e}=\mu_{n}-\mu_{p},
\end{aligned}
\end{equation}
and the charge neutrality condition is fulfilled by:
\begin{equation}
\sum_{B}q_B\rho_B+\sum_{D}q_D\rho_D-\rho_K-\rho_e-\rho_{\mu}=0
\end{equation}
The chemical potential of baryons, $\Delta$ and leptons read:
\begin{equation}
\mu_{i}=\sqrt{k_{F}^{i 2}+m_{i}^{* 2}}+g_{\omega i}\omega+g_{\phi i}\phi+g_{\rho i}\tau_{3 i} \rho, i=B,D
\end{equation}
\begin{equation}
\mu_{l}=\sqrt{k_{F}^{l 2}+m_{l}^{2}},
\end{equation}
where $k_{F}^{i}$ is the Fermi momentum and the $m_{i}^{*}$ is the effective mass of baryon and $\Delta$ resonances, which can be related to the scalar meson field as $m_{i}^{*}=m_{i}-g_{\sigma i}\sigma$, and $k_{F}^{l}$ is the Fermi momentum of the lepton $l$($\mu$,e).
\par
The total energy density of the system with Kaon condensation reads then $\varepsilon=\varepsilon_{(B,D)}+\varepsilon_{K}$, where $\varepsilon_{B,D}$ is the energy density of baryons and $\Delta$ resonances, can be given as
\begin{equation}
\begin{aligned}
\varepsilon_{B,D}=&\sum_{i=B,D}\frac{\gamma}{(2\pi)^{3}}\int_{0}^{k_{Fi}}\sqrt{m_{i}^{*}+k^{2}}d^{3}k+\frac{1}{2}m_{\omega}^{2}\omega^{2}\\
&+\frac{\xi}{8}g_{\omega N}^{4}\omega^{4}+\frac{1}{2}m_{\sigma}^{2}\sigma^{2}+\frac{\kappa}{6}g_{\sigma N}^{3}\sigma^{3}+\frac{\lambda}{24}g_{\sigma N}^{4}\sigma^{4}\\
&+\frac{1}{2}m_{\rho}^{2}\rho^{2}+3\Lambda_{\nu}g_{\rho N}^{2}g_{\omega N}^{2}\omega^{2}\rho^{2}+\frac{1}{2}m_{\phi}^{2}\phi^{2}\\
&+\frac{1}{\pi^{2}}\sum_{l}\int_{0}^{k_{Fl}}\sqrt{k^{2}+m_{l}^{2}}k^{2}dk,
\label{eq6}
\end{aligned}
\end{equation}
And the energy contributed by the Kaon condensation $\varepsilon_{K}$ is
\begin{equation}
\varepsilon_{K}=2m_{K}^{* 2}K^{*}K=m_{K}^{*}\rho_{K},
\end{equation}
The Kaon does not contribute directly to the pressure as it is a (s-wave) Bose condensate so that the expression of pressure reads
\begin{equation}
P=\sum_{i=B,D}\mu_{i}\rho_{i}+\sum_{l=\mu,e}\mu_{l}\rho_{l}-\varepsilon,
\end{equation}
With the obtained $\varepsilon$ and $P$, the mass-radius relation and other relevant quantities of neutron star can be obtained by solving the Oppenheimer and Volkoff equation \cite{Baldo:1997ag}.
\begin{equation}
\begin{aligned}
\frac{dP(r)}{dr}=&-\frac{GM(r)\varepsilon}{r^{2}}(1+\frac{P}{\varepsilon C^{2}})(1+\frac{4\pi r^{3}P}{M(r)C^{2}})\\
&\times(1-\frac{2GM(r)}{rC^{2}})^{-1},
\end{aligned}
\end{equation}
\begin{equation}
dM(r)=4\pi r^{2}\varepsilon(r)dr
\end{equation}
The tidal deformability of a neutron star is reduced as a dimensionless form \cite{hinderer2008tidal,Postnikov:2010yn}.
\begin{equation}
\Lambda=\frac{2}{3}k_{2}C^{-5}
\end{equation}
where $C=GM/R$, the second Love number $k_{2}$ can be fixed
simultaneously with the structures of compact stars \cite{Hinderer:2009ca}.
\par
The $\sigma$-cut scheme \cite{Maslov:2015lma}, which is able to stiffen the EOS above saturation density, adds in the original Lagrangian density, the function \cite{Maslov:2015lma,Song:2011jh,Kolomeitsev:2015qia}
\begin{equation}
\Delta U(\sigma)=\alpha \ln(1+exp[\beta(f-f_{s,core})])
\end{equation}
where $f=g_{\sigma N}{\sigma}/M_{N}$ and $f_{s,core}=f_{0}+c_{\sigma}(1-f_{0})$. $M_{N}$ is the nucleon mass. $f_{0}$ is the value of $f$ at saturation density, equal to 0.31 for the IUFSU model. $c_{\sigma}$ is a positive parameter that we can adjust. The smaller $c_{\sigma}$ is, the stronger the effect of the $\sigma$-cut scheme becomes. However, we must be careful that this scheme would not affect the saturation properties of nuclear matter, in our previous work we discussed in detail the choice of parameter $c_{\sigma}$ \cite{Ma:2022fmu}. In this paper, we take $c_{\sigma}$=0.15 to satisfy the maximum mass constraint. $\alpha$ and $\beta$ are constants, taken to be $4.822 \times 10^{-4}M_{N}^{4}$ and 120 as in Ref. \cite{Maslov:2015lma}. This scheme stiffens the EOS by quenching the decreasing of the effective mass of the nucleon $M_{N}^{*}=M_{N}(1-f)$ at high density.
\par
\section{\label{sec:level3}Results}
\begin{figure}
\includegraphics{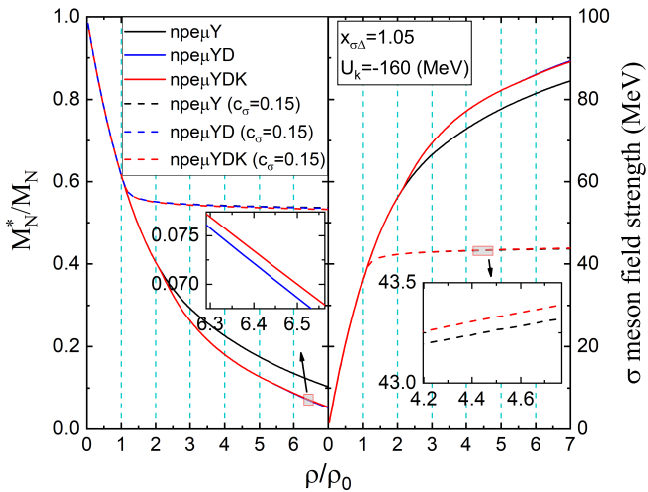}
\caption{Effective mass and $\sigma$ meson strength of nucleons versus baryon density in NS matter using and not using $\sigma$-cut scheme.}
\label{fig1}
\end{figure}
First, we studied the effect of the $\sigma$-cut scheme on IUFSU model. In Fig. \ref{fig1}, we plot the ratio of the effective mass of nucleons to the rest mass and the $\sigma$ meson strength  as a function of the baryon density, where $\rho_{0}$ is the saturation density, and we choose $x_{\sigma \Delta}$=1.05 and $U_{K} = -160 $ MeV to consider $\Delta$ resonance and Kaon condensation. From the left panel, we can see that when $\rho \leq \rho_{0}$, the effect mass is almost same as nucleons-only matter and unchanged by the $\sigma$-cut scheme, when $\rho$\textgreater$\rho_{0}$, the effect mass dropped to around $0.55 M_{N}$. And it is  obviously observed that   under  the $\sigma$-cut scheme  considering $\Delta$ and $K^{-}$ in the EOS or not has   very tiny effect on the effective mass  of nucleons.   From right panel, when $\rho$ \textgreater $\rho_0$, the $\sigma$ meson field
strength is quenched at high baryon density, this is what we want by using the $\sigma$-cut scheme.
\begin{figure}
\centering
\includegraphics{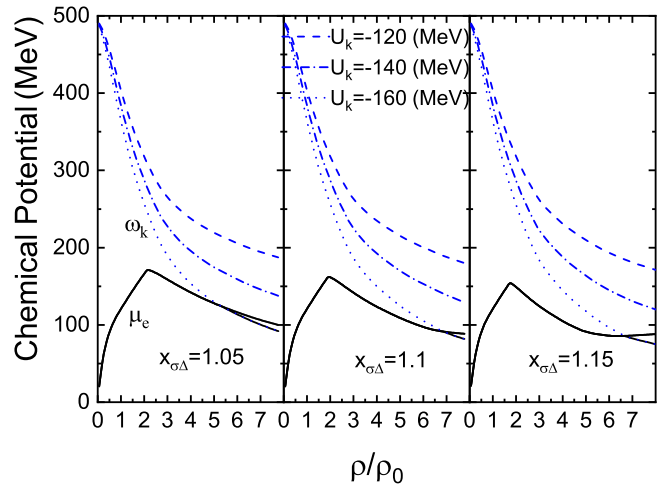}
\caption{Kaon meson energy($\omega_{K}$) and electronic chemical potential($\mu_{e}$) as a function of $\rho_B$ with different $x_{\sigma \Delta}$ and $U_K$, and without $\sigma$-cut scheme.}
\label{fig2}
\end{figure}
\par
\begin{figure}
\centering
\includegraphics{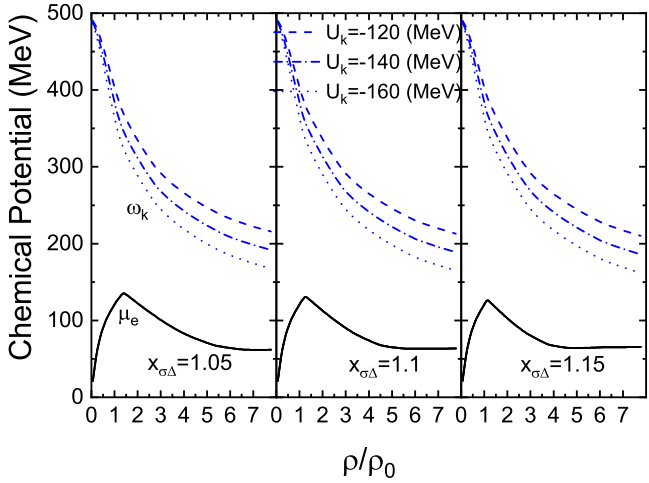}
\caption{Kaon energy($\omega_{k}$) and electron chemical potential($u_{e}$) as a function of baryon density with different $x_{\sigma \Delta}$ and $U_K$, $c_{\sigma}$=0.15}
\label{fig3}
\end{figure}

In Fig. \ref{fig2}, we plot the chemical potential of $K^{-}$ and $e^{-}$ as a function of baryon density. With the increase of density, the energy $\omega_{K}$ of a test Kaon in the pure normal phase can be computed as a function of the nucleon density. The Kaon energy($\omega_K$) will decrease while the electron chemical potential ($\mu_{e}$) increases with the density. When the condition $\omega_{K}=\mu_{e}$ is achieved, the Kaon will occupy a small fraction of the total volume. We can see that both $x_{\sigma \Delta}$ and $U_{K}$  affect the Kaon meson condensation. From $x_{\sigma \Delta}$ = 1.05 to 1.15, there is no intersection  between $\omega_{K}$ and $\mu_{e}$ when $U_k = -120$ MeV and -140 MeV. 
The intersection of $\omega_{K}$ and $\mu_{e}$ is only possible when $U_{K}$=-160MeV, which means that the smaller the optical potential of the $K^-$ at saturated nuclear matter is, the greater the possibility of the Koan condensation is.  When we choose $\sigma$-cut scheme(Fig. \ref{fig3}), there is no intersection between $\omega_{K}$ and $\mu_{e}$.
The decrease of $\sigma$ meson field strength slows down the decline of $\omega_{K}$, and makes the appearance of $K^{-}$ difficult. We list the threshold densities $n_{cr}$ for Kaon condensation for different values of $K^{-}$ optical potential depths $U_{K}$ in Table \ref{tab:Table 5.}.
\begin{table}
\caption{\label{tab:Table 5.}
Threshold densities $n_{cr}$ (in units of $\rho$/$\rho_0$) for Kaon condensation in dense nuclear matter for different values of $K^{-}$ optical potential depths $U_{K}$ (in units of MeV) without $\sigma\mbox{-}$cut scheme.
}
\begin{ruledtabular}
\begin{tabular}{c|ccc}
\multirow{2}{*}{$U_{K}$ (MeV)} & \multicolumn{2}{c}{$n_{cr}(K^{-})$} \\
 & \multicolumn{1}{c}{$x_{\sigma \Delta}=1.05$} & $x_{\sigma \Delta}=1.1$ & $x_{\sigma \Delta}=1.15$ \\ \hline
-120 & \multicolumn{1}{c}{$none$} & $none$ & $none$ \\
-140 & \multicolumn{1}{c}{$none$} & $none$ & $none$ \\
-160 & \multicolumn{1}{c}{5.24} & 6.79 & 6.73 \\
\end{tabular}
\end{ruledtabular}
\end{table}
\begin{figure}
\centering
\includegraphics{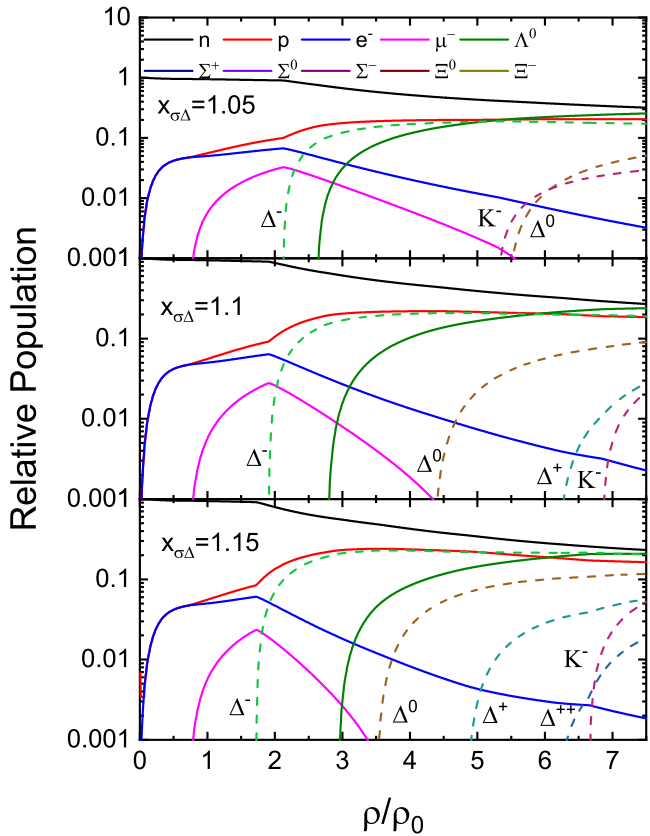}
\caption{Relative population of particles versus baryon density without $\sigma$-cut scheme with $x_{\sigma \Delta}$=1.05, $x_{\sigma \Delta}$=1.1, $x_{\sigma \Delta}$=1.15 and $K^{-}$ potential depth of $U_{K}=-160$ MeV, dashed lines denote $K^{-}$ and $\Delta$ resonance.}
\label{fig4}
\end{figure}
\par
Fig. \ref{fig4} shows the relative population of particles versus baryon density with $x_{\sigma \Delta}$=1.05, $x_{\sigma \Delta}$=1,1 and $x_{\sigma \Delta}$=1.15, $U_{K}$=-160MeV. We find that as $x_{\sigma \Delta}$ increases, the critical density of $\Lambda^{0}$ moves to a higher density region, while the as of density of leptons moves to a lower density, and in particular, as $\mu^{-}$ disappear, $\Delta^{0}$ starts to appear and the critical density of $\Delta$ resonance moves to a lower density region. When $x_{\sigma \Delta}=1.1$, $\Delta^{+}$ appears at $6.21\rho_{0}$, and for $x_{\sigma \Delta}=1.15$, $\Delta^{++}$ appears at $6.14\rho_0$ while the critical density of $K^{-}$ meson occurs at $6.79\rho_0$ and $6.73\rho_0$, respectively. 
\par
\begin{figure}
\centering
\includegraphics{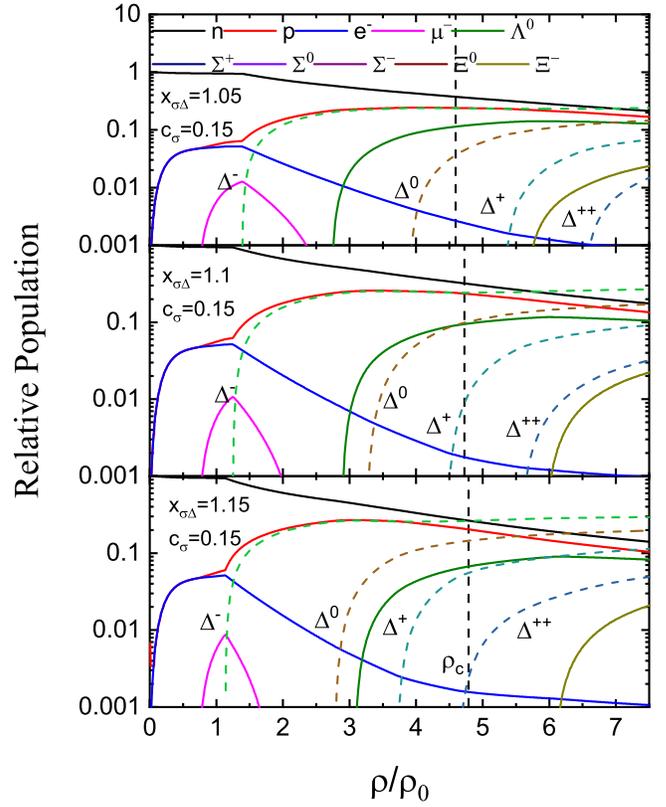}
\caption{Relative population of particles versus baryon density with $\sigma$-cut scheme($c_{\sigma}$=0.15), $x_{\sigma \Delta}$=1.05, $x_{\sigma \Delta}$=1.1, $x_{\sigma \Delta}$=1.15 and $K^{-}$ potential depth of $U_{K}=-160$ MeV, dashed lines denote $K^{-}$ and $\Delta$ resonance.}
\label{fig5}
\end{figure}

Next we examine the effect of the $\sigma$-cut scheme on the particle population. This is plotted in the Fig. \ref{fig5}. From Fig. \ref{fig3}, 
we determined that no $K^{-}$ is generated when using $\sigma$-cut scheme, as there is no intersection between $\omega_{K}$ and $\mu_{e}$, 
but the $\Delta$ resonance has some interesting variations. The $K^{-}$ disappear, and as the decrease of $\mu^{-}$, the $\Delta^{+}$ and $\Delta^{++}$ increase as the charge balance conditions lead to the appearance of new hyperons $\Xi^{-}$, suggesting that hyperons are more favorable as neutralizers of positive charges compared to leptons. As $x_{\sigma \Delta}$ increases from 1.05 to 1.15, the critical value of $\Delta$ resonance shifts to lower density and the central energy density($\rho_c$) will move towards the high density area, in particular, when $x_{\sigma \Delta}=1.15$, the critical density of $\Delta^{0}$ moves before the $\Lambda^{0}$. From these figures, 
it can be concluded that the appearance of Kaon meson condensation is more likely to suppress the hyperon production than the $\Delta$ resonance. Although the $\sigma$-cut scheme leads to the disappearance of Kaon meson condensation, it does not change the relationship between the $\Delta$ resonance as $x_{\sigma \Delta}$ varies.
\begin{figure}
\includegraphics{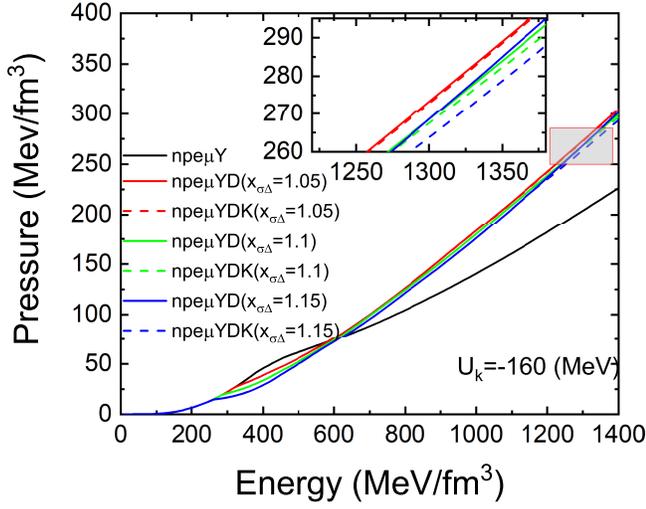}
\caption{Pressure versus energy density without the $\sigma$-cut scheme. The solid line is for n, p, leptons and hyperons whereas others are with additional $\Delta$ resonance, dashed lines contain $K^{-}$ mesons and exhibit $U_{K}=-160$ MeV.}
\label{fig6}
\end{figure}
\par
Then we can discuss some properties of the neutron star. Fig. \ref{fig6} shows pressure as a function of energy density in NS matter containing $\Delta$ resonance and $K^{-}$ without $\sigma$-cut scheme. We can see from the enlarged area in the figure that the addition of $K^{-}$ softens the equation of state to some extent, although this is not particularly significant with the onset of the $\Delta$ resonance. As $x_{\omega \Delta}$ increases, the EOS will get softer, eventually leading to a decrease in the maximum mass of the neutron star. It is worth mentioning that $
\Delta$ resonance softens the EOS when the energy density is between $300MeV/fm^3$ and $600MeV/fm^3$ and stiffens significantly \textgreater $600MeV/fm^{3}$ compared to the case where only hyperons are included, and intensifies with increasing $x_{\sigma \Delta}$(1.05$\rightarrow$1.15), which suggests the existence of a softer EOS in the low-density region, and the recent constraints on tidal deformation and radius point to this. 

\begin{figure}
\centering
\includegraphics{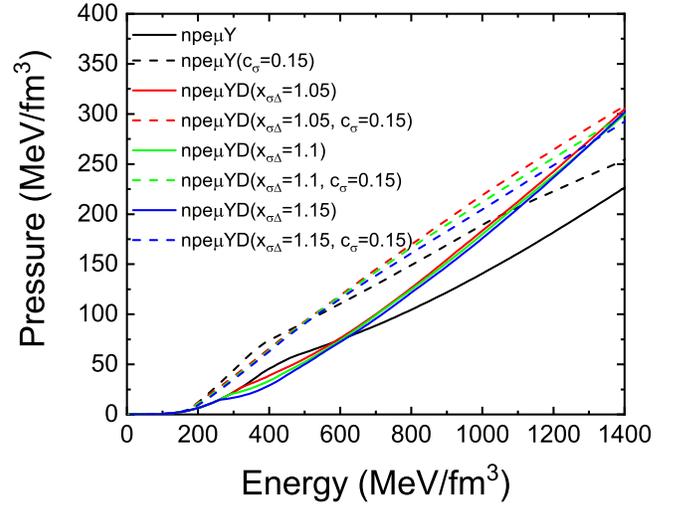}
\caption{Pressure versus energy density with $\sigma$-cut scheme. The solid line is for n, p, leptons and hyperons whereas others are with additional $\Delta$ resonance, dashed line exhibits $c_{\sigma}$=0.15.}
\label{fig7}
\end{figure}
When $\sigma$-cut scheme is considered, we plot the EOS in Fig. \ref{fig7}. From Fig. \ref{fig3} we determined that there is no $K^{-}$ when using $\sigma$-cut scheme, so the composition contains only hyperons and $\Delta$, we can see that $\sigma$-cut scheme significantly stiffens the EOS, and it is the truncated intensity of $\sigma$ meson field strength in Fig .\ref{fig1} that leads to this result, and still retains the EOS softening feature in the low density region. The EOS obtained by this way can generate heavier $2M_{\odot}$ NS by solving the TOV equation, in order to eliminate "hyperon puzzle".
\begin{figure}
\centering
\includegraphics{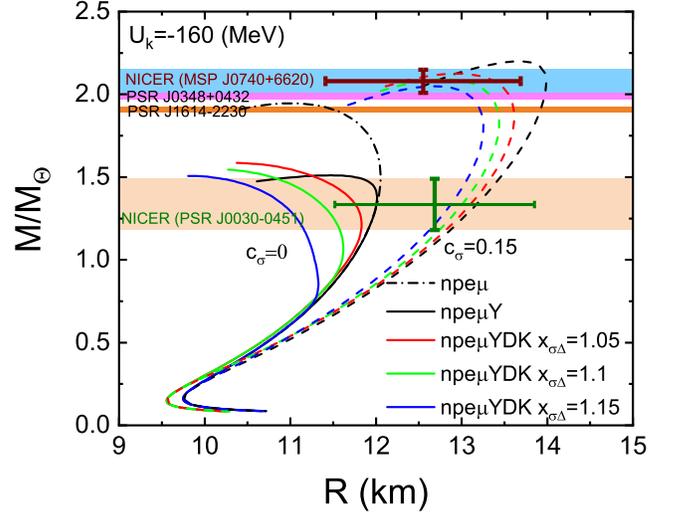}
\caption{Mass-radius relation using and not using $\sigma$-cut scheme in NS matter including hyperons, $\Delta$ resonance and $K^{-}$. The solid lines denoted without $\sigma$-cut, dashed lines denoted $c_\sigma$=0.15. The horizontal bars indicate the observational constraints of PSR J1614 - 2230 \cite{Demorest:2010bx,Ozel:2010bz,NANOGrav:2017wvv,Fonseca:2016tux}, PSR J0348 + 0432 \cite{antoniadis2013massive}, MSP J0740 + 6620 \cite{fonseca2021refined} and PSR J0030-0451 \cite{riley2019nicer}.}
\label{fig8}
\end{figure}
\par
The results of mass-radius relation for NS discussed here and shown in Fig. \ref{fig8}. The constraints from the observables of massive neutron stars, PSR J1614-2230 \cite{Demorest:2010bx,Ozel:2010bz,NANOGrav:2017wvv,Fonseca:2016tux}and PSR J034+0432 \cite{antoniadis2013massive} are also shown as the shaded bands. The Neutron star Interior Composition Explorer (NICER) collaboration reported an accurate
measurement of mass and radius of PSR J0030+0451 \cite{riley2019nicer} in 2019, and MSP J0740 + 6620 in 2021 \cite{fonseca2021refined}. For the solid lines without $\sigma$-cut, different coupling parameters $x_{\sigma \Delta}$ have a significant effect on the maximum mass and radius of NS, it shows that the $\Delta$ resonance increases the maximum mass of NS and decreases the radius. As the increase of $x_{\sigma \Delta}$(1.05$\rightarrow$1.15), the maximum mass decreases, but is still greater than in the case of pure hyperons. The dashed lines denote $c_{\sigma}$=0.15, this scheme can significantly increase the maximum mass of the neutron star and make it heavier than $2M_\odot$, also accords with the constraints from gravitational wave and NICER(MSP J0740+6620). Note that there is no appearance of $K^{-}$ when $c_{\sigma}$=0.15 from Fig. \ref{fig3}. We list the simultaneous measurement of radius for MSP J0740 + 6620  and PSR J0030 - 0451  by the NICER data and maximum mass of the neutron star for various values of $x_{\sigma \Delta}$ in Table \ref{Table6.}.
\par
\begin{table*}[ht]
\caption{\label{Table6.}
The maximum mass (in unit of solar mass $M_{\odot}$) and radius(km) in NS matter including hyperons, $\Delta$ resonance and $K^{-}$ using and not using $\sigma$-cut scheme with potential $U_{K}$=-160MeV.}
\begin{ruledtabular}
\begin{tabular}{c|ccc|ccc|cc|cc}
\multirow{2}{*}{} & \multicolumn{2}{c}{without ${\sigma}$-cut} & & \multicolumn{2}{c}{$c_{\sigma}$=0.15} & & \multicolumn{2}{c|}{MSP J0740+6620 \cite{fonseca2021refined}} & \multicolumn{2}{c}{PSR J0030-0451 \cite{riley2021nicer}} \\
 & \multicolumn{1}{c}{M} & ${\rho_c}$ & R & \multicolumn{1}{c}{M} & ${\rho_c}$ & R & \multicolumn{1}{c}{M} & R & \multicolumn{1}{c}{M} & R \\ \hline
 ($n,p$) & \multicolumn{1}{c}{1.93} & 1.029 & 11.14 & \multicolumn{1}{c}{-} & - & - & \multicolumn{1}{c}{} &  & \multicolumn{1}{c}{} &  \\
($n,p,Y$) & \multicolumn{1}{c}{1.51} & 0.87 & 11.47 & \multicolumn{1}{c}{2.2} & 0.58 & 13.65 & \multicolumn{1}{c}{\multirow{5}{*}{$2.08\pm0.07$}} & \multirow{5}{*}{$12.39^{+1.3}_{-0.98}$} & \multicolumn{1}{c}{\multirow{5}{*}{$1.34^{+0.15}_{-0.16}$}} & \multirow{5}{*}{$12.71^{+1.14}_{-1.19}$} \\
$x_{\sigma \Delta}=1.05$($n,p,Y,D,(K^{-})$) & \multicolumn{1}{c}{1.58} & 1.24 & 10.37 & \multicolumn{1}{c}{2.12} & 0.71 & 12.93 & \multicolumn{1}{c}{} &  & \multicolumn{1}{c}{} &  \\
$x_{\sigma \Delta}=1.1$($n,p,Y,D,(K^{-})$) & \multicolumn{1}{c}{1.54} & 1.3 & 10.27 & \multicolumn{1}{c}{2.09} & 0.73 & 12.84 & \multicolumn{1}{c}{} &  & \multicolumn{1}{c}{} &  \\
$x_{\sigma \Delta}=1.15$($n,p,Y,D,(K^{-})$) & \multicolumn{1}{c}{1.51} & 1.37 & 9.9 & \multicolumn{1}{c}{2.05} & 0.74 & 12.64 & \multicolumn{1}{c}{} &  & \multicolumn{1}{c}{} &  \\
\end{tabular}
\end{ruledtabular}
\end{table*}
\begin{figure}
\centering
\includegraphics{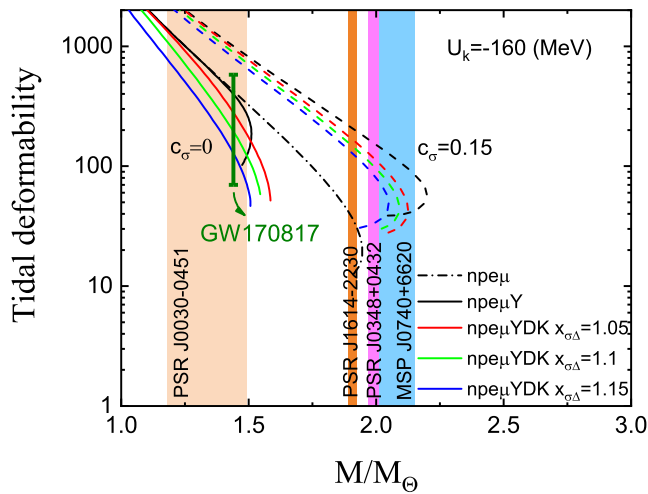}
\caption{The dimensionless tidal deformability as a function of star mass. The solid line indicates without $\sigma$-cut scheme, the dashed line indicates that $c_{\sigma}$=0.15. And the constraints from GW170817 event for tidal deformability is shown.}
\label{fig9}
\end{figure}
\par
The tidal deformability $\Lambda$, as a function of neutron stars is shown in Fig. \ref{fig9}. From the gravitation wave of BNS merger in GW170817, it was extracted as $\Lambda_{1.4}=190^{+390}_{-120}$ at 1.4$M_{\odot}$ \cite{LIGOScientific:2018cki}. From the figure we can see that the $\sigma$-cut scheme with the stiffer EOS has the larger $\Lambda_{1.4}$ and heavier masses, whose $\Lambda_{1.4}$ are out the constraint of GW170817, while the softer EOS satisfies the constraints of GW170817 and has smaller radii without $\sigma$-cut scheme, and the $\Lambda$ is still within the bound of GW170817 after considering the $\Delta$ resonance. With the strong constraint on the compositions of compact stars by the observational tidal deformability, we think it is necessary to consider $\Delta$ resonance in the softer EOS in the event of GW170817. And the tidal deformability of the neutron star at 2.0$M_{\odot}$, which is expected to be measured in the future gravitational wave events from the binary neutron-star merger.

\section{\label{sec:level4}Summary}

In this paper, we have discussed the $\Delta$ resonance and Kaon meson condensation inside the neutron star under the IUFSU model, due to the recent rapid results of astronomical observations on the radii and tidal deformations of compact stars. However, the maximum masses of neutron stars generated by the softer EOS(hyperon puzzle) cannot approach 2.0$M_{\odot}$, which did not satisfy the constraints from the massive neutron star observables, so we take $\sigma$-cut scheme and got the maximum mass heavier than 2$M_{\odot}$. 
\par{}
We find that the Kaon condensation cannot appear in the hyperons and $\Delta$ resonance with our parameter $U_{K}$=-120MeV and -140MeV, it occurs only at $U_{K}$=-160MeV, and the ${\Delta}$ resonance also shifts the Kaon meson toward the high-density region. In NS matter containing hyperons and $\Delta$ resonances, the effect of Kaon meson on EOS is very insignificant.
\par{}
On the other hand, we investigated the effect of $x_{\sigma \Delta}$ on the $\Delta$ resonance, for the $\Delta$ coupling constants, we take $x_{\sigma \Delta}$=1.05, 1.1 and 1.15, and the value of  $x_{\sigma \Delta}$ has great influence 
on the relative population of particles as a function of the baryon density.
We find that the inclusion of $\Delta$ resonance shifts the critical density of hyperons towards the high density region from $x_{\sigma \Delta}=1.05$ to $x_{\sigma \Delta}=1.15$ and the critical density of $\Delta$ resonance will move toward the low density region. Also, EOS softens as $x_{\sigma \Delta}$ increases.
\par{}  
When the $\sigma$-cut scheme is not used, we find that the softer EOS considering $\Delta$ resonance is still within the $\Lambda_{1.4}$ range of the GW170817, and this may suggest the existence of a softer EOS in the low-density region, while the softer EOS satisfies the constraints of GW170817 and has smaller radii. When we used $\sigma$-cut and take the parameter $c_{\sigma}=0.15$, we find that the maximum mass and radius of NSs obtained under this model are close to the NICER(MSP J0740+6620) constraint. For tidal deformability of neutron stars with maximum mass above $2M_{\odot}$, future gravitational wave events of binary neutron star mergers may provide new constraints.

\bibliography{2022-11-15-delta}

\bibliographystyle{unsrt}

\end{document}